\documentclass[%
reprint,
superscriptaddress,
nofootinbib,
amsmath,amssymb,
aps,
prd,
floatfix,
]{revtex4}

\usepackage{color}
\usepackage{graphicx}
\usepackage{dcolumn}
\usepackage{bm}
\usepackage{hyperref}
\usepackage[normalem]{ulem}

\hypersetup{
    colorlinks=true,
    linkcolor=blue,
    citecolor=blue,
    filecolor=black,
    urlcolor=blue,
}

\renewcommand{\eqref}[1]{Eq.~\ref{#1}}

\newcommand{\figref}[1]{Fig.~\ref{#1}}

\newcommand{\mpl}{\ensuremath{M_\text{P}}} 
\newcommand{\sclt}{\ensuremath{a_\perp}} 
\newcommand{\sclp}{\ensuremath{a_\parallel}} 

\newcommand{\mtime}{\ensuremath{H_{\text{I}}}}
\newcommand{\mfld}{\mpl}
\newcommand{\scl}{\ensuremath{a}} 
\newcommand{\hub}{\ensuremath{H}} 
\newcommand{\grd}{\ensuremath{\tilde{G}}}
\newcommand{\clEst}{\ensuremath{\hat{C}_2}}
\newcommand{\clObs}{\ensuremath{\hat{C}_2^{\text{obs}}}}

\newcommand{\clObsV}{\ensuremath{253.6\, \mu\mathrm{K}^2}}
\newcommand{\clThV}{\ensuremath{1124.1\, \mu\mathrm{K}^2}}
\newcommand{\zetaE}{\ensuremath{\zeta_{\text{end}}}}

\def\|{{ \, || \,}}

\begin{document}

\title{Constraining cosmological ultra-large scale structure using numerical relativity}

\author{Jonathan Braden}
  \email{j.braden@ucl.ac.uk}
  \affiliation{Department of Physics and Astronomy, University College London, London WC1E 6BT, U.K.}

\author{Matthew C. Johnson}
  \email{mjohnson@perimeterinstitute.ca}
  \affiliation{Department of Physics and Astronomy, York University, Toronto, Ontario, M3J 1P3, Canada}
  \affiliation{Perimeter Institute for Theoretical Physics, Waterloo, Ontario N2L 2Y5, Canada}

\author{Hiranya V. Peiris}
  \email{h.peiris@ucl.ac.uk}
  \affiliation{Department of Physics and Astronomy, University College London, London WC1E 6BT, U.K.}
\affiliation{The Oskar Klein Centre, Department of Physics, AlbaNova, Stockholm University, SE-106 91 Stockholm, Sweden}

\author{Anthony Aguirre}
  \email{aguirre@scipp.ucsc.edu}
  \affiliation{SCIPP and Department of Physics, University of California, Santa Cruz, CA, 95064, USA}

\date{\today}

\begin{abstract}
  Cosmic inflation, a period of accelerated expansion in the early universe, can give rise to large amplitude ultra-large scale inhomogeneities on distance scales comparable to or larger than the observable universe. The cosmic microwave background (CMB) anisotropy on the largest angular scales is sensitive to such inhomogeneities and can be used to constrain the presence of ultra-large scale structure (ULSS). We numerically evolve nonlinear inhomogeneities present at the beginning of inflation in full general relativity to assess the CMB quadrupole constraint on the amplitude of the initial fluctuations and the size of the observable universe relative to a length scale characterizing the ULSS. To obtain a statistically meaningful ensemble of simulations, we adopt a toy model in which inhomogeneities are injected along a preferred direction. We compute the likelihood function for the CMB quadrupole including both ULSS and the standard quantum fluctuations produced during inflation. We compute the posterior given the observed CMB quadrupole, finding that when including gravitational nonlinearities, ULSS curvature perturbations of order unity are allowed by the data, even on length scales not too much larger than the size of the observable universe. To demonstrate the robustness of our conclusions, we also explore a semi-analytic model for the ULSS which reproduces our numerical results for the case of planar symmetry, and which can be extended to ULSS with a three-dimensional inhomogeneity structure.  Our results illustrate the utility and importance of numerical relativity for constraining early universe cosmology.
\end{abstract}

\maketitle


\section{Introduction}

Cosmic inflation, a postulated era of accelerated expansion in the early universe, has become an integral part of modern cosmology; but while inflation provides a dynamical mechanism to produce the initial conditions for the standard hot Big Bang cosmology, the initial conditions for inflation itself are far more uncertain. 
Cosmological measurements are bounded by the cosmological horizon, placing fundamental limits on our direct knowledge of the universe on arbitrarily large scales. However, many inflationary models --- for example, those including ``eternal" inflation~\cite{Aguirre:2007gy,Guth:2007ng} or starting from inhomogeneous initial conditions --- lead to a rich and complicated structure on ultra-large scales compared to our own local Hubble volume. It is thus of great interest to look for ways to probe this structure using ``local" measurements of the observable universe. 

In this work we study the effects of nonlinear inhomogeneities present at the beginning of single-scalar field inflation on the quadrupole of the cosmic microwave background (CMB) radiation observed today, a phenomenon known as the Grischuk-Zel'dovich (GZ) effect~\cite{GZeffect}. Because pre-inflationary physics primarily affects the largest scales of the observable universe, the CMB quadrupole is the most important observable in this context. Previous studies of the GZ effect (e.g., Refs.~\cite{Castro:2003bk,Turner:1991dn}) neglect the impact of gravitational nonlinearities, basing their inferences on linear cosmological perturbation theory and Gaussian statistics. In this context, the expected imprint of primordial inhomogeneities on the CMB quadrupole shrinks exponentially with an increasing number of inflationary $e$-folds. However, the large gravitational nonlinearities studied in this scenario could plausibly lead to significant non-Gaussianity and departures from linear perturbation theory, which in turn affect our inferences about ultra-large scale structure (ULSS) from the CMB quadrupole. While it is true that even in the highly non-linear regime not all realizations of inflation will give rise to observable ULSS (e.g. because some regions undergo many $e$-folds of inflation), our primary goal is to sharpen the connection between pre-inflationary physics and cosmological observables to determine precisely which scenarios can be ruled out from observation.

The potential importance of gravitational nonlinearities motivates the fully general relativistic numerical treatment that we undertake in this work.
For computational efficiency, we assume primordial inhomogeneities are only present along a preferred spatial direction, allowing for one-dimensional simulations.
The model is specified by the inflationary potential, and by the spectral shape and amplitude of primordial inhomogeneities in the inflaton field.
Using exquisitely accurate numerical techniques, we evolve realizations of these pre-inflationary fluctuations well into the post-inflationary regime using mere seconds of computing time, while maintaining convergence to the level of machine precision. This numerical efficiency allows us to sample many realizations of the initial conditions to build up a statistical description of observables in our model.

For three qualitatively different choices of the inflationary potential, we vary the initial fluctuation amplitude and build up a set of probability distributions over the locally observed $\ell = 2$ CMB multipole, $a_{20}$, at different spatial positions.  We find that independent of the inflationary model, and under a variety of physically plausible weighting schemes for spatial positions, nonlinear gravitational effects yield a highly non-Gaussian distribution for $a_{20}$. 
We then fold into our calculation the standard smaller-wavelength vacuum fluctuations generated by inflation,
to obtain the probability distribution of power in the observed CMB quadrupole. The relative contribution of the ULSS and vacuum fluctuations is controlled by a variable that maps primordial length scales to present-day length scales. 

We compute the full posterior over the initial fluctuation amplitude and mapping parameter, comparing with what would have been obtained if gravitational nonlinearities had not been taken into account. We  conclude it is essential to accurately model gravitational nonlinearities to make accurate inferences about ULSS from measurements of the CMB quadrupole. Based on a semi-analytic extrapolation of our results, we expect similar qualitative conclusions to hold in the case where no symmetry assumptions are made, and we comment on the broader implications of our result to this more realistic scenario. More generally, our results motivate a systematic study of the influence of strong gravity in the early universe on cosmological observables.

Our results fit within the broader context of numerical relativity as a cosmological tool~\cite{Goldwirth:1989pr,Goldwirth:1990pm,KurkiSuonio:1993fg,Johnson:2011wt,Wainwright:2013lea,East:2015ggf,Giblin:2015vwq,Bentivegna:2015flc,Mertens:2015ttp,Adamek:2015eda,Kleban:2016sqm,Adamek:2016zes,Clough:2016ymm}. While much of this previous work focused on the relativistic dynamics of either the pre-inflationary or matter-dominated Universe, this work is among the first to self-consistently connect initial conditions to cosmological observables.

\section{Models and numerical methods}
We consider $\alpha$ attractor models of inflation~\cite{Kallosh:2013yoa}, with Lagrangian density
\begin{equation}\label{eq:lagrangian}
  \mathcal{L} = -\frac{1}{2}\partial_\mu\phi\partial^\mu\phi - V_0\left(1-e^{-\sqrt{\frac{2}{3\alpha}}\frac{\phi}{M_P}} \right)^2 \, .
\end{equation}
Large- and small-field models can be related to the primordial tensor-to-scalar ratio through the Lyth bound~\cite{Lyth:1996im}.
The $\alpha$ attractor models conveniently interpolate between large-field inflation at $\alpha \gg 1$ and small-field inflation at $\alpha \ll 1$.  Current limits favor intermediate and small-field models~\cite{Ade:2015tva}. We analyzed three representative models: $\alpha \rightarrow \infty$ (equivalent to a quadratic inflationary potential~\cite{Linde:1983gd}), $\alpha = 2/3$ (similar to Starobinsky's $R^2$ inflation~\cite{Starobinsky:1980te}, where $\alpha=1$), and $\alpha = 2/300$ (where inflation occurs on a flat plateau).  In all three cases we found qualitatively similar behavior.  To illustrate the basic phenomenology, we present results for  $\alpha=2/300$ below.

We assume our spacetime possesses two spatial translation isometries and choose a synchronous gauge with metric
\begin{equation}\label{eq:metric}
ds^2 = -d\tau^2 + \sclp^2(x,\tau)dx^2 + \sclt^2(x,\tau)\left(dy^2+dz^2\right) \, .
\end{equation}
The full set of evolution and constraint equations resulting from Einstein's equations are presented in Appendix~\ref{app:numerical}.
We introduce the field momenta $\Pi_\phi \equiv \sclp \dot{\phi}$ (with $\dot{\phi} \equiv \partial_\tau \phi$) and the extrinsic curvatures ${{K^x}_x}$ and ${{K^y}_y}$.
We also define $H_{\text{I}}^2 \equiv V(\bar{\phi}(\tau=0)) / 3 \mpl^2$, where $\bar{\phi}(\tau=0)$ is the average value of the field on the initial conditions surface $\tau = 0$.
Below, we measure fields in terms of \mfld. 

\begin{figure*}
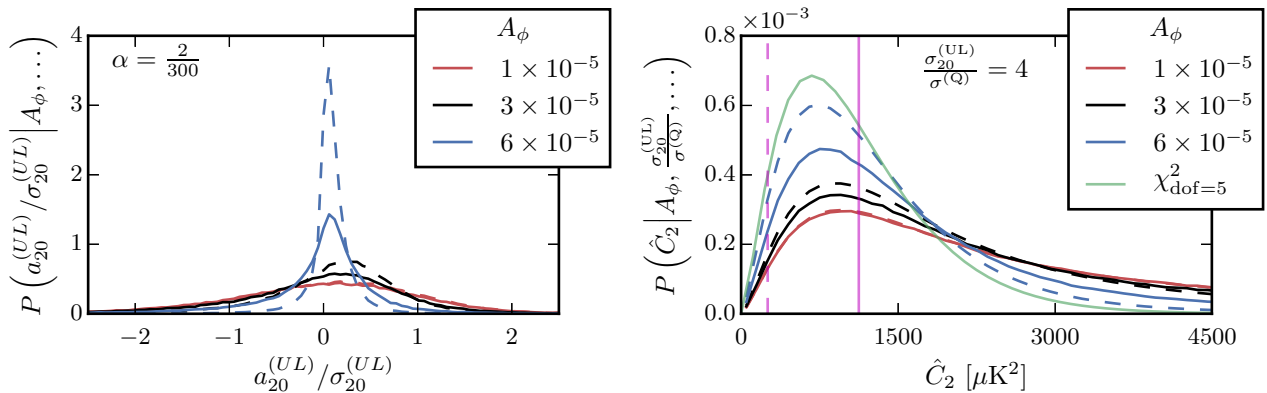

  \includegraphics[width=3.375in]{{{pdfs_a20_vary_weights}}}
  \includegraphics[width=3.375in]{{{c2_mu10_withweights_varyamp_units}}}
  \caption{\textit{Left}: PDF for the $\ell =2$ CMB multipole $\hat{a}_{20}^{\rm (UL)}$ generated by ULSS, scaled to $\sigma_{20}^{\rm (UL)} \equiv \sqrt{\left\langle \left(\hat{a}_{20}^{\rm (UL)}\right)^2 \right\rangle}$.  As the amplitude of the initial fluctuations is varied, the distribution becomes increasingly non-Gaussian and peaked.  \textit{Right}:  Sampling distribution of the estimator $\clEst$ in \eqref{eq:C2realization}.  The solid vertical magenta line is the mean value predicted by the best-fit $\Lambda$CDM model from the {\it Planck} data, and the dashed vertical magenta line is the observed value. In both panels, solid lines are the distributions for comoving volume weighting and dashed lines are for physical volume weighting.}
  \label{fig:quadrupoleplots}
\end{figure*}

We choose the coordinates $x, y,$ and $z$ to measure proper distance on the initial spatial slice, corresponding to ${\sclp(\tau=0,x)} = {\sclt(\tau=0,x)} = 1$. The initial conditions for each simulation can then be specified entirely in terms of $\phi$ and $\Pi_\phi$. Once these are given, we solve the momentum constraint for ${{K^y}_y}$, and finally substitute into the Hamiltonian constraint to obtain an algebraic equation for ${{K^x}_x}$.

We take the scalar field on the initial surface to be ${\phi (\tau=0,x) = \bar{\phi} + \delta\hat{\phi}(x)}$ and ${\Pi_\phi(x) = \bar{\Pi}_\phi = 0}$. The mean field $\bar{\phi}$ is set to obtain $60$  $e$-folds of inflation in the homogeneous slow-roll approximation. The normalization $V_0$ is set to match the amplitude of scalar power measured by the {\it Planck} satellite~\cite{Ade:2015xua}.  
The fluctuations $\delta\hat{\phi}(x)$ are drawn from a one-dimensional Gaussian random field with a band-pass filtered white noise spectrum
\begin{equation}
  \delta\hat{\phi}(x_i) = A_\phi \sum_{n=n_{\mathrm{IR}}}^{n_{\mathrm{UV}}}e^{i k_n x_i}\grd_n \qquad i=1,\dots,N \, ,
  \label{eqn:phiFluc}
\end{equation}
where $k_n \equiv 2\pi \frac{n}{L}$, $x_i = idx$, $\grd_n$ are realizations of complex Gaussian random deviates with $\langle |\grd|^2 \rangle = 2$, and $A_\phi$ is a free parameter.
We choose $n_{\text{IR}}=1$, $n_{\text{UV}} = \sqrt{3}\mtime L$ and a box size $\mtime L= 256/\sqrt{3}$, corresponding to modes spanning the range $\frac{2\pi}{256}\sqrt{3} < \mtime^{-1} k < 2\pi\sqrt{3}$ in wavenumber. Note that we do not model the pre-inflationary phase that gives rise to these initial conditions.

The evolution equations are solved numerically from the initial pre-inflationary hypersurface through several post-inflationary $e$-folds of expansion.
Using the numerical techniques described in Appendix~\ref{app:numerical}, we are able to achieve machine-precision accuracy with run times on the order of seconds.
We use this numerical efficiency to explore four orders of magnitude in $A_\phi$, running 100 realizations of the Gaussian initial conditions at each amplitude.

\section{Extracting cosmological observables}
The CMB quadrupole as viewed from each location in the simulation is determined by the comoving curvature perturbation $\zetaE$ at the end of inflation, along with the cosmological redshift to the end-of-inflation hypersurface. There are a variety of methods to find $\zetaE$~\cite{Wainwright:2013lea,Xue:2013bva,Johnson:2015gma}. Here we follow the $\delta\mathcal{N}$ formalism~\cite{Salopek:1990jq,Sasaki:1995aw} and compute 
\begin{equation}
\zeta(H) = \frac{1}{6}\ln\mathrm{det}\gamma_{ij}|_{\sclp=\sclt=1}^{-\frac{1}{3}{K_i}^i = H} = \ln\left(a\right) \, ,
\end{equation}
where ${\gamma_{ij} = \mathrm{diag}(\sclp^2,\sclt^2,\sclt^2)}$ is the spatial three-metric on fixed $\tau$ slices and ${a \equiv \left(\sclp\sclt^2\right)^{1/3}}$.
$\zeta$ is defined on hypersurfaces of constant {$\hub \equiv -\frac{1}{3}K_i^i$} as measured by comoving observers.
For notational convenience, we also define
\begin{equation}
  \zeta_{\parallel,\perp}(H) = \ln(a_{\parallel,\perp})|_{\sclp=\sclt=1}^{-\frac{1}{3}{K_i}^i = H}
\end{equation}
evaluated on the same hypersurfaces.
It also proves convenient to consider only the spatial fluctuations in $\zeta$, so we further define
\begin{equation}
  \label{eqn:del-zeta}
  \delta\zeta \equiv \zeta - \bar{\zeta}
\end{equation}
with analogous definitions for $\delta\zeta_\parallel$ and $\delta\zeta_\perp$.
$\bar{\zeta} = \langle\zeta\rangle$ is an average taken over the constant $H$ surface with each grid site weighted equally.
For the adiabatic perturbations we study here, $\delta\zeta$ freezes out within a few $e$-folds of the start of our simulations, so that $\bar{\zeta}$ is a function of $H$ alone and $\delta\zeta$ is a function of ${\bf x}$ alone.
The squared proper distance along these hypersurfaces is
\begin{equation}
  dx_{H=\mathrm{const}}^2 = \sclp^2\left(1 - \frac{H'^2}{\sclp^2\dot{H}^2} \right) dx^2 + a_\perp^2\left(dy^2+dz^2\right) \, ,
\end{equation}
where $H' \equiv \partial_x H$ is evaluated with respect to synchonous spatial coordinates at fixed $\tau$. The hypersurface on which inflation ends is defined by the first occurrence of {$\epsilon_H \equiv -d\ln\hub / d\ln\scl = 1$} at each spatial position.
After a short transient, $\epsilon_H$ is a function of $H$ alone. 
Therefore, $\zetaE$ corresponds to the particular choice $\zeta(H_{\text{end}})$. Note that $\zeta$ quantifies the overall expansion of a local packet of geodesics, so variations in the value of $\zeta$ as a function of position encodes the comoving curvature perturbation.
 
As outlined in Appendix~\ref{app:loc}, the Taylor series coefficients of $\zetaE$ around each point ${\bf x}_0$ along the end-of-inflation hypersurface are related to the properties of a set of locally-defined nearly homogeneous patches, which can in turn be related to the observed CMB at each position. 
The constant term in the expansion $\zetaE ({\bf x}_0)$ is the natural logarithm of the local scale factor,
which is not directly observable. The first derivative term also produces no observable signatures, since pure gradients in the gravitational potential are pure gauge in linear perturbation theory~\cite{Turner:1991dn,Erickcek:2008jp}. The leading observable is therefore the quadratic term in the expansion, which maps onto a CMB quadrupole through the Sachs-Wolfe and Integrated Sachs-Wolfe effect. For the situation we focus on, this truncated expansion encapsulates all observable effects of ULSS.

In appropriate coordinates, only the $m=0$ mode of the $\ell = 2$ CMB temperature anisotropy is non-zero.  Along a fixed $H$ slice, the physical size of the last scattering surface $R_{\rm ls}$ is independent of position, so it is convenient to approximate $a_{20}$ by evaluating $\zeta$ at the physical position of the last scattering surface for each position ${\bf x}_0$
\begin{equation}\label{eq:a20}
a_{20}^{\rm (UL)} \simeq -F T_{\rm CMB}(L_{\rm obs}H_{\rm I})^2 e^{2\bar{\zeta}}\partial_{x_p}^2 \zeta \simeq -F T_{\rm CMB}(L_{\rm obs} H_{\rm I})^2 e^{-(\zeta_\parallel-\bar{\zeta})} \frac{\partial}{\partial\bar{x}} \left[e^{-(\zeta_\parallel-\bar{\zeta})}\frac{\partial\zeta}{\partial\bar{x}} \right] \, . 
\end{equation}
The derivatives are taken with respect to dimensionless coordinates $\bar{x} = H_{\rm I}x$.
Here $F$ is a constant dimensionless factor encoding order unity projection coefficients, a correction from both the early and late time Integrated Sachs-Wolfe effect, and a conversion from proper distance along the end-of-inflation hypersurface to comoving distance at last scattering. $L_{\rm obs}=e^{-\bar{\zeta}}R_{\rm ls}$ is the comoving radius of the surface of last scattering, and $H_{\rm I}$ is defined above. The final approximation assumes $(H'/\sclp\dot{H})^2 \ll 1$, which is valid when the local Hubble expansion rate varies only on scales much larger than the local Hubble horizon. 

Due to uncertainties in the post-inflation expansion history which depend on the details of (p)reheating, the combination $L_{\rm obs}H_{\rm I}$ is a free parameter.
We have chosen our conventions such that in the homogeneous limit, $L_{\rm obs} H_{\rm I} = 1$ is the necessary condition for solving the horizon problem, which translates into the observable universe being descended from a single causally-connected domain at the beginning of inflation of physical size $H_{\rm I}^{-1}$.
In the small-amplitude limit, $H_{\rm I}^{-1}$ translates into the local comoving scale of inhomogeneity.
For the case of strong inhomogeneity, this identification is lost due to the strong spatial variations of the local scale factor on the end-of-inflation hypersurface.
 
\section{Simulation Results}
For each simulation, we extract the locally-observed CMB quadrupole~\eqref{eq:a20} at each grid site from $\zetaE$.
Additionally, we must assign a probability (or weight) to an imaginary observer at each grid point.
We consider two weighting schemes: comoving volume weighting where each grid point is weighted equally, and physical volume weighting where each grid point is weighted by the total growth in physical volume during inflation ($e^{3\zetaE(x)} / \langle e^{3\zetaE(x)}\rangle$).
Because each coordinate position $x$ labels a geodesic, our simulation coordinates naturally preserve comoving volume (see also Ref.~\cite{Aguirre:2006ak}).
Empirical PDFs are created by binning the $a_{20}^{\rm (UL)}$ samples by their assigned weights.

Several representative PDFs for the small-field model of inflation ($\alpha = 2/300$) are shown in the left panel of \figref{fig:quadrupoleplots}. The PDFs for the comoving volume weighting scheme are denoted by solid lines and those for physical volume weighting by dashed lines. Note that the PDFs in \figref{fig:quadrupoleplots} are normalized to their variance; un-normalized distributions would have a variance that increases with $A_\phi$, spanning ${8\times 10^{-2} < \sigma_{20}^{\rm (UL)} \lesssim 10^{13}}$ for $10^{-6} \leq A_\phi \leq 10^{-3}$.  

As expected, for small $A_\phi$ the distribution is Gaussian.
In this regime, there is no significant difference between the weighting schemes.
As $A_\phi$ is increased, gravitational nonlinearities become more important, with variations in $\zeta$ becoming order unity for ${A_\phi \agt 2\times 10^{-4}}$ in the small-field model.
In this same limit, the PDFs gain a more pronounced non-Gaussian shape, primarily in the form of an increasing kurtosis.
The PDFs also develop a sharp maximum at positive values for $a_{20}^{\rm (UL)}$, indicating a bias towards peaks in $\zeta$ over troughs, as well as a non-zero mean.
The emergence of a peak in the distribution offset from the origin and the breaking of reflection symmetry about the origin arises primarily from our assumption that the ULSS is planar symmetric, which breaks statistical isotropy.  This can be seen explicitly by comparing~\figref{fig:quadrupoleplots} and~\figref{fig:3D-distributions}.
For very large amplitudes, the distribution becomes very sharply peaked about $a_{20}^{\rm (UL)}=0$.

In the highly nonlinear regime, most of the volume of the end-of-inflation surface descends from regions which have experienced a larger amount of inflation than in the homogeneous limit (see also Refs.~\cite{Garriga:2005av,Linde:2006nw}).
This explains the emergence of a peak at positive values in the $a_{20}^{\rm (UL)}$ PDF, as regions that inflate more are biased to be peaks rather than troughs.\footnote{Note that this correlation between negative Laplacian and positive $a_{20}$ is a consequence of our planar symmetry assumption}, which breaks statistical isotropy. In the statistically isotropic case, it is the spatial curvature that develops a positive valued peak the same reason; see Sec.~\ref{sec:3d}. It also explains the limiting form of the PDF, since for most of the volume on the reheating surface, inhomogeneities would have been blown up to extremely large scales.

As a consequence, both weighting schemes produce similar results as illustrated in \figref{fig:quadrupoleplots}.
The primary difference is that the amplitude $A_\phi$ at which non-Gaussianities become apparent is lower for the volume weighting scheme.
Similarly, this is also the main difference between choices of inflationary potentials, where the threshold $A_\phi$ at which the $a_{20}^{\rm (UL)}$ distribution becomes non-Gaussian increases as $\alpha$ is increased.  This is simply a reflection of the increased sensitivity to initial condtions of small-field models compared to large-field models~\cite{Brustein:1992nk}.

To facilitate a comparison of the predictions of our toy model with the observed CMB angular power spectrum, we combine the effects on $a_{20}$ of ULSS and the standard inflationary quantum fluctuations that give rise to structure on sub-horizon scales.
The observed CMB quadrupole is constructed from the sample variance $\clEst$.
In our toy model, a particular realization yields
\begin{equation}\label{eq:C2realization}
\hat{C}_2 = \frac{1}{5} \left[ \left( a_{20}^{\rm (UL)} + a_{20}^{\rm (Q)} \right)^2 +  \sum_{m=-2, \ m \neq 0}^2 \left(a^{\rm (Q)}_{2m}\right)^2 \right] \, ,
\end{equation}
where $a_{20}^{\rm (UL)}$ is drawn from distributions analogous to those shown in the left panel of Fig.~\ref{fig:quadrupoleplots}, and each $a_{2m}^{\rm (Q)}$ is drawn from a Gaussian of variance $\left(\sigma^{\rm (Q)}\right)^2 = \clThV$, in agreement with the current best-fit $\Lambda$CDM parameters from the {\it Planck} satellite~\cite{Ade:2015xua}.
A more realistic scenario would have ULSS contributions to all 5 of the $a_{2 m}$ in addition to a contribution to the locally-observed spatial curvature; we explore a toy model where this is the case below.

In the right panel of Fig.~\ref{fig:quadrupoleplots}, we show the predicted probability distribution over $\clEst$, generated by drawing realizations $\clEst$ (\eqref{eq:C2realization}) while fixing the variance of $a_{20}^{\rm (UL)}$ to be 4 times larger than that of $a_{20}^{\rm (Q)}$. For comparison, we show the PDF over $\clEst$ in the absence of ULSS (a $\chi^2$ distribution with 5 degrees of freedom).  We will comment briefly on this in Section~\ref{sec:3d}.

Surprisingly, Fig.~\ref{fig:quadrupoleplots} shows that {\em for fixed relative RMS}, larger initial amplitudes $A_\phi$ become increasingly indistinguishable from the case of no ULSS contribution. At the level of Eq.~\ref{eq:C2realization}, this is because the width of the peak in the PDF is much smaller than the variance of the distribution, and nearly all $a_{20}^{\rm (UL)}$ draws come from the vicinity of the peak very near zero.

\section{Semianalytic Approach to ULSS Quadrupole}
Before turning to the observational constraints associated with the ULSS quadrupole, we first outline a semi-analytic approximation for the distribution of the temperature quadrupole $a_{20}^{\rm (UL)}$ in the ULSS. This explicitly demonstrates the role of the local volume expansion in determining the qualitative behaviour of the distribution (and thus the lack of constraining power for large initial field fluctuation amplitudes). Furthermore, the approach can be extended to the full three-dimensional problem without relying on a (computationally infeasible) Monte Carlo sampling of field configurations using three-dimensional numerical relativity simulations. A derivation of the appropriate mapping between the locally observed quadrupole moments/spatial curvature and the curvature perturbation in the case with no assumption of planar symmetry is presented in Appendix~\ref{app:loc}.

Identifying $\sclp \simeq \sclt \simeq e^{\zeta({\bf x}_0)}$, the general result~\eqref{eqn:a2m-full} agrees with~\eqref{eq:a20} to leading order in $\Delta\zeta$ as defined in Appendix~\eqref{eqn:local-zeta-pert}, yielding
\begin{equation}
  \label{eqn:a20-planar}
  a_{2m}^{\rm (UL)}({\bf x}_0) \propto -\left(H_{\rm I}L_{\rm obs}\right)^2 e^{-2\zeta({\bf x}_0)} \zeta''({\bf x}_0) \, \delta_{m,0} \, .
\end{equation}
Here, ${}'$ represents a spatial derivative with respect to the spatial coordinates along a fixed Hubble slice. Within this approximation, the goal of numerical simulations is reduced to the determination of joint one-point probability distributions for the local values of $\zeta$ and $\zeta''$.  As is clear from the derivation in the Appendix, the reduction from the full field statistics to a few one-point distributions occurs because of the superhorizon nature of the perturbations.

\begin{figure}
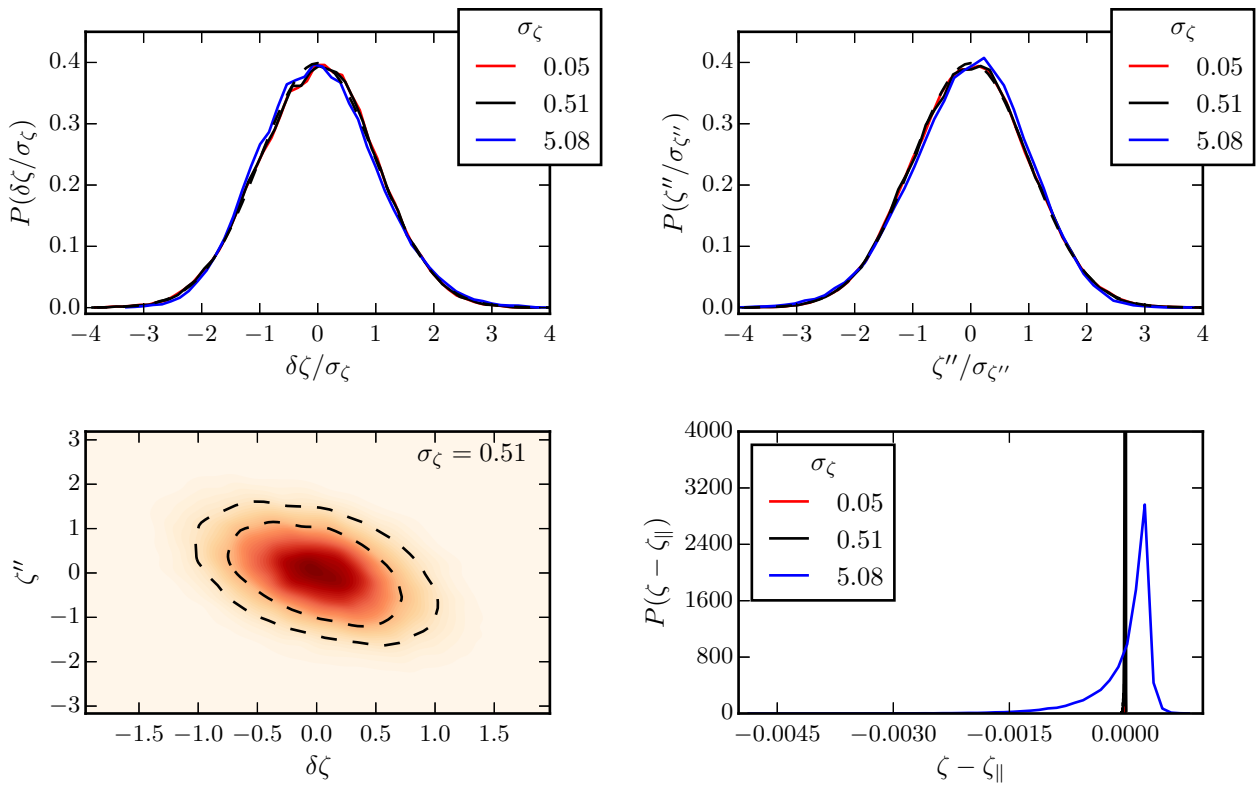

  \includegraphics[width=3.375in]{{{zeta-dist-varya}}}
  \includegraphics[width=3.375in]{{{zetapp-dist-varya}}} \\
  \includegraphics[width=3.375in]{{{zeta-zetapp-correlation}}}
  \includegraphics[width=3.375in]{{{zeta-diff-pdf}}}
  \caption{\emph{Top Row:} Normalized one-point distributions of $\delta\zeta$ and $\zeta''$ using comoving volume weighting and derivatives with respect to synchronous coordinates on fixed $H$ slices.  The different coloured curves correspond to $A_\phi = 10^{-5}$ (\emph{red}), $10^{-4}$ (\emph{black}), and $10^{-3}$ (\emph{blue}), with corresponding $\sigma_\zeta$ as labelled in the figures.  For reference, the black dashed line is a Gaussian PDF.
    \emph{Bottom Row:} The left panel shows the joint PDF of $\delta\zeta$ and $\partial_{xx}\zeta$ for $A_\phi = 10^{-4}$, with the two black lines showing the numerically approximated countours at $e^{-1}$ and $e^{-2}$ of the maximal value. The right panel shows the PDF of $\zeta-\zeta_\parallel$, demonstrating the tight correlation between the two expansion factors and also validating the approximation $\bar{\gamma}_{ij} \approx \delta_{ij}$.}
  \label{fig:one-point-stats}
\end{figure}

In~\figref{fig:one-point-stats} we show normalized distributions of these two quantities as extracted from our numerical simulations, assuming comoving volume weighting.
Remarkably, we observe that the one-point distributions of $\zeta$ and its derivatives are well-approximated by Gaussians, even in the non-linear regime ($\sigma_\zeta \gtrsim 1$).  Furthermore, $(\zeta_\parallel-\zeta)$, which measures the anisotropic expansion (equivalently, the deviation of $\bar{\gamma}_{ij}$ from the identity matrix), is small throughout the simulation volume.
Since we have
\begin{equation}
  \gamma_{ij} = e^{2\zeta}\mathrm{diag}\left(\begin{array}{ccc} e^{-2(\zeta-\zeta_\parallel)}, & e^{(\zeta-\zeta_\parallel)}, & e^{(\zeta-\zeta_\parallel)} \end{array}\right)
\end{equation}
this justifies our approximation $\sclp \simeq \sclt \simeq e^{\zeta({\bf x}_0)}$ above so that we can take $\bar{\gamma}_{ij} \approx \delta_{ij}$ as in Appendix~\ref{app:loc}.

\begin{figure}
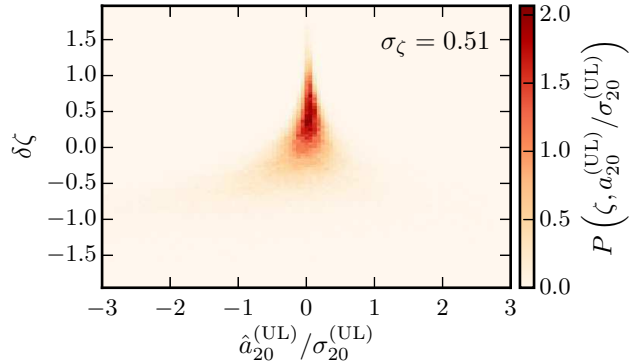

  \includegraphics[width=3.375in]{{{a20-zeta-correlation}}}
    \caption{The joint probability distribution of $a_{20}^{\rm (UL)}$ and $\delta\zeta$ extracted from numerical simulations with $A_\phi = 10^{-4}$.  We have chosen an initial field fluctuation amplitude $A_\phi$ just probing the regime of $\mathcal{O}(1)$ $\zeta$ fluctuations.  As is evident, the emergence of the peak in $a_{20}^{\rm (UL)}$ around zero is associated with large positive fluctuations in $\zeta$. Physically, these regions expanded more, so the local scale of the ULSS is larger than in the nearly homogeneous limit.}
\label{fig:a20-zeta-correlation}
\end{figure}

As shown in~\figref{fig:a20-zeta-correlation}, the emergence of a large peak around the origin in $\hat{a}_{20}^{\rm (UL)}$ is associated with large positive fluctuations in $\zeta$, which act to increase the local scale of the ULSS relative to the physical size of the last scattering surface. This effect is evident from the exponential prefactor in~\eqref{eqn:a20-planar}. Further, if instead of comoving volume weighting, we weight by the total volume expansion from the initial condition hypersurface, the effect will be to upweight the values of $\hat{a}_{20}^{\rm (UL)}$ in the peak around the origin, resulting in the emergence of the peak at smaller values of $A_\phi$, exactly as we observed previously.
The above observations suggest we model $\zeta({\bf x})$ as a Gaussian random field, and furthermore approximate $\zeta_\parallel = \zeta$.
To the order of approximation used in this section, $\hat{a}_{20}^{\rm (UL)}$ is a function of two correlated Gaussian random deviates, whose joint probability is given by
\begin{equation}
  P({\bf G}) = \frac{1}{\sqrt{(2\pi)^2\mathrm{det}C_G}}e^{-\frac{1}{2}{\bf G}^TC_G^{-1}{\bf G}}\, ,
\end{equation}
where we have defined ${\bf G} = \left[\begin{array}{cc}\delta\zeta & \zeta''\end{array} \right]^T$.
The covariance matrix $C_G$ is given by
\begin{equation}
  C_G \equiv \left[
  \begin{array}{cc}
    \langle \delta\zeta^2 \rangle  & \langle\zeta\zeta'' \rangle \\
    \langle \zeta\zeta''\rangle & \langle {\zeta''}^2\rangle
  \end{array} \right]
  = 
  \left[
  \begin{array}{cc}
    \langle \delta\zeta^2 \rangle & -\langle {\zeta'}^2\rangle \\
    -\langle {\zeta'}^2 \rangle & \langle {\zeta''}^2\rangle
  \end{array} \right]
  \equiv
  \sigma_0^2 \left[ \begin{array}{cc}
    1                 & -\bar{\sigma}_1^2 \\
    -\bar{\sigma}_1^2 &  \bar{\sigma}_2^2
  \end{array} \right] \, .
\end{equation}
In the above, $\langle\cdot\rangle$ represents an average with uniform weighting of grid points and $\delta\zeta \equiv \zeta - \langle\zeta\rangle$.
For convenience, we have defined normalized spectral moments
\begin{equation}
  \label{eqn:shape-params}
  \bar{\sigma}_i^2 = \frac{\sigma_i^2}{\sigma_0^2} = \frac{\langle(\partial^{2i}\zeta)^2\rangle}{\langle \delta\zeta^2\rangle} = \frac{\int dk (k^2)^i P(k)}{\int dk P(k)} \, ,
\end{equation}
where $P(k) = \langle|\tilde{\zeta}_k|^2\rangle$ is the power spectrum of $\zeta(x)$ in synchronous spatial coordinates on fixed $H$ slices.  The $\bar{\sigma}_i$'s parametrize the shape of the power spectrum, while $\sigma_0$ modulates the amplitude of the $\zeta$ fluctuations.  The determination of the distribution of $\hat{a}_{20}^{\rm (UL)}$ has thus been reduced to a determination of three numbers.
In the more general planar symmetric case we have $a_\parallel \neq a_\perp \neq a$, leading to a nontrivial unit determinant spatial 3-metric $\bar{\gamma}_{ij}$.  To full second order in spatial derivatives, $a_{20}^{\rm (UL)}$ becomes a function of six correlated random deviates $\zeta,\zeta',\zeta'',\zeta_\parallel$, $\zeta_\parallel'$, and $\zeta_\parallel''$.  For example, assuming $\zeta_\parallel = \zeta$, the approximation used in~\eqref{eq:a20} also requires information about $\zeta'$.  We will not pursue this semi-analytic extension here, since we verified that for the simulations presented in this paper, including additional nonlinear corrections induced only small changes in the empirically derived PDFs for $a_{20}^{\rm (UL)}$.

\begin{figure}[!h]
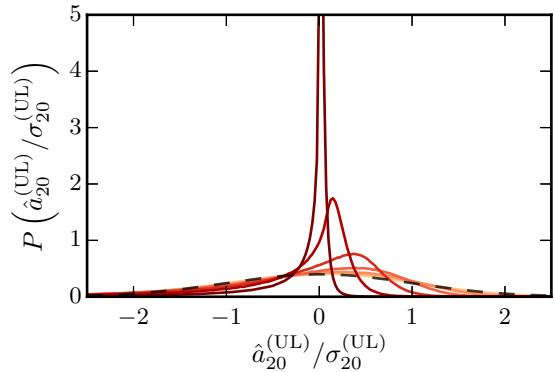

  \includegraphics[width=3.375in]{{{a20_GRF_vary-sig0}}}
  \caption{Distributions of $a_{20}^{\rm (UL)}$ in the Gaussian approximation for $\zeta$ as we vary $\sigma_\zeta = \sqrt{\langle\delta\zeta^2\rangle}$ at fixed $\bar{\sigma}_1^2=1$ and $\bar{\sigma}_2^2=1.56$ (see~\eqref{eqn:shape-params}).  For ease of visualization, the distributions are normalized to $\sigma_{20}^{\rm (UL)} \equiv \sqrt{\left\langle\left(\hat{a}_{20}^{\rm (UL)}\right)^2\right\rangle} \propto \sigma_\zeta e^{4\sigma_\zeta^2}\sqrt{\left(\bar{\sigma}_2^2 + 16\sigma_\zeta^2\bar{\sigma}_1^4\right)}$. This corresponds to fixing the spectrum of the Gaussian random field $\zeta(x)$ while adjusting it's overall amplitude to increase the variance of $\zeta$.  For reference, the black dashed line is an appropriately scaled Gaussian distribution for $a_{20}^{\rm (UL)}$.}
  \label{fig:a20-gaussian-vary-params}
\end{figure}

Figure~\ref{fig:a20-gaussian-vary-params} illustrates the $a_{20}^{\rm (UL)}$ distributions predicted by the simple Gaussian model for the ULSS.  For illustration, we have fixed the shape parameters $\bar{\sigma}_1$ and $\bar{\sigma}_2$ to constants, while adjusting the overall amplitude of the spectrum $P(k)$.  We verified that for the range of $A_\phi$ and $k_{\rm cut}$ used in our simulations, this is a reasonable approximation for the dynamically generated ULSS $\zeta$ fields.  

\section{Observational constraints}
The CMB quadrupole obtained using the \texttt{Commander} approach\footnote{Available from the Planck Legacy Archive at \url{http://pla.esac.esa.int/pla/\#cosmology}.} on {\it Planck} 2015 data is $\clObs = \clObsV$. The central value lies near the lower 95\% confidence level bound of the best-fit $\Lambda$CDM model in the absence of ULSS. This arguably anomalously low value for the CMB quadrupole has received note since the first full-sky measurements of the CMB angular power spectrum by the COBE satellite~\cite{1996ApJ...464L..17H}. Our model does not alleviate this issue. 

\begin{figure}
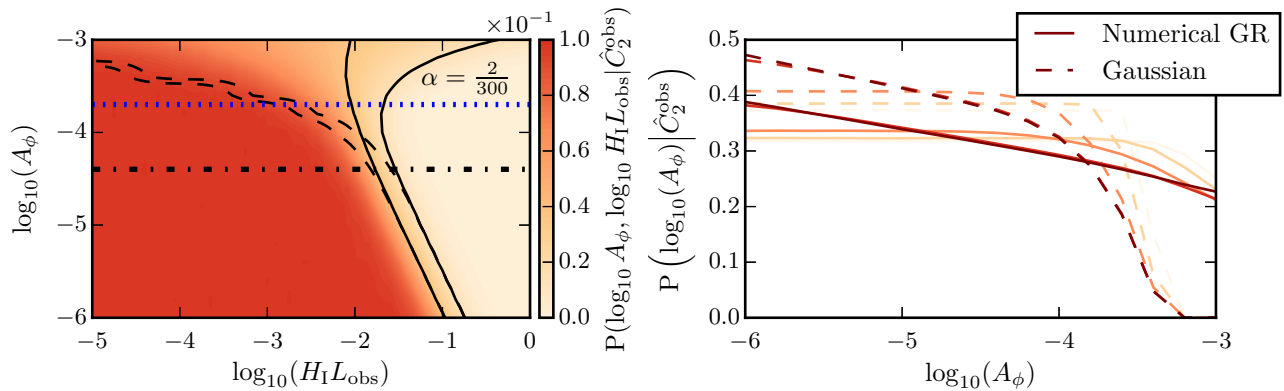

   \includegraphics[width=3.375in]{{{posterior_c2_xls_vs_Aphi}}}
   \includegraphics[width=3.375in]{{{prob_aphi_marginalised_scan}}}
   \caption{\emph{Left}: The posterior probability distribution for our model parameters $\log_{10}(A_\phi)$ and $\log_{10}(H_{\rm I}L_{\rm obs})$ with uniform priors in the range $[-6,-3]$ and $[-5,0]$ respectively.  The solid lines are the contours at which the posterior drops to $e^{-1}$ and $e^{-2}$ of its maximal value.  The dashed curves are the same contours if we model $a^{\rm (UL)}_{20}$ in~\eqref{eq:C2realization} as a Gaussian random variable with the same RMS as the distributions in~\figref{fig:quadrupoleplots}.  The blue dotted line indicates the initial amplitude $A_\phi$ at which $\sigma_\zeta = 1$.
     The black dot-dashed line indicates the point at which $\left\langle e^{-4\delta\zeta}\left(\frac{\partial^2\zeta}{\partial (H_Ix)^2}\right)^2\right\rangle=1$, which acts as a proxy for when the distribution of $a_{20}$ begins to become non-Gaussian.
     \emph{Right}: The marginalized posterior for the initial scalar field fluctuation amplitude $A_\phi$ obtained by marginalizing over $\log_{10}\left(H_{\rm I}L_{\rm obs}\right)$ for both the numerically generated (\emph{solid}) and Gaussian (\emph{dashed}) $a_{20}^{\rm (UL)}$ distributions.  Clearly, correctly accounting for the inhomogeneous local expansion induced by gravitational nonlinearities completely changes the nature of our inferences at large $A_\phi$.  To illustrate the robustness of this conclusion to our prior choice, we show curves for a range of prior widths.  Each curve shows the marginalized posterior assuming $\log_{10}(H_{\rm I}L_{\rm obs})$ is uniformly distributed in the range $[-5,\log_{10}(L_{\rm cut})]$ with varying choices of $\log_{10}L_{\rm cut} \in \{-4,-3,-2,-1,0\}$.  Darker shades correspond to larger values of $\log_{10}L_{\rm cut}$.}
  \label{fig:exclusionplot}
\end{figure}

Rather, given $\clObs$ we derive constraints on the presence of ULSS by computing the posterior ${\rm P} (\log_{10}A_\phi, \log_{10} L_{\rm obs} H_{\rm I} | \hat{C}_2^{\rm obs})$, shown in~\figref{fig:exclusionplot} for the model with $\alpha = 2/300$ and comoving volume weighting.
For the purposes of numerical evaluation, it is convenient to rewrite~\eqref{eq:C2realization} as
  \begin{equation}
    \hat{C}_2 = \left(\sigma^{\rm (Q)}\right)^2\left[ \left(R\frac{\hat{a}_{20}^{\rm (UL)}}{\sigma_{20}^{\rm (UL)}} + \hat{\mathcal{N}}_0\right)^2 + \hat{\chi}^2_{{\rm dof}=4}\right]
  \end{equation}
  where $\hat{\mathcal{N}}$ is drawn from a Gaussian distribution of unit variance, and $\hat{\chi}^2_{{\rm dof}=4}$ from a chi-squared distribution with four degrees of freedom.
  The ratio of fluctuation amplitudes in the superhorizon versus subhorizon stucture is given by
  \begin{equation}
    R^2 = \frac{\langle (a_{20}^{\rm (UL)})^2\rangle}{\left(\sigma^{\rm (Q)}\right)^2} = F^2\left(\frac{T_{\rm CMB}}{\sigma^{\rm (Q)}}\right)^2\left(H_{\rm I}L_{\rm obs}\right)^{4}\langle (\bar{a}^{\rm (UL)}_{20})^2\rangle 
  \end{equation}
  where $\bar{a}_{20}^{\rm (UL)} = e^{-2\left(\zeta_\parallel-\bar{\zeta}\right)}\left(\zeta'' - \zeta_\parallel'\zeta'\right)$.
  In this decomposition, the value of $A_\phi$ enters into our final posterior both through the shape of $a_{20}^{\rm (UL)}$ as encoded in $a_{20}^{\rm (UL)}/\sigma_{20}^{\rm (UL)}$ (such as those shown in~\figref{fig:quadrupoleplots}) and through the relation between $R$ and $\left(H_{\rm I}L_{\rm obs}\right)^2$ whose $A_\phi$ dependence is carred by $\langle(\bar{a}_{2m}^{\rm (UL)})^2\rangle$.

In computing the posterior, we assume uniform priors on $\log_{10}A_\phi$ and $\log_{10}H_{\rm I}L_{\rm obs}$ in the ranges $-6 \leq \log_{10}A_\phi \leq -3$ and $-5 \leq \log_{10}L_{\rm obs} H_{\rm I}  \leq 0$. The upper bound on $\log_{10}A_\phi$ corresponds to the point where our numerics break down (well into the nonlinear regime), and the upper bound on $\log_{10}H_{\rm I}L_{\rm obs}$ is the value required to solve the horizon problem. Note that the model without ULSS is at $A_\phi = L_{\rm obs} H_{\rm I}  = 0$, which maximizes the posterior.
  
In Fig.~\ref{fig:exclusionplot}, we also show contours (dashed lines) for the posterior generated under the assumption that the ultra-large scale $a_{20}^{(\rm UL)}$  are drawn from a Gaussian distribution of equal variance to the numerically derived non-Gaussian PDFs. This is an extrapolation of Gaussian fluctuations into the non-linear regime.
As expected, for small $A_\phi$ where the PDFs are approximately Gaussian, there is good agreement.
However, the posterior has far more weight than the Gaussian case when $A_\phi \gtrsim 10^{-4.4}$, where $\zeta$ fluctuations are order unity.
As a result, when considering the marginalized posterior for the initial fluctuation amplitude $A_\phi$, the cutoff that appears at large $A_\phi$ in the Gaussian case disappears as illustrated in the right panel of~\figref{fig:exclusionplot}.
Therefore, it is relatively more {\em difficult} to constrain ULSS in the nonlinear regime.

Comparing the posterior generated for the different models of inflation and different weighting schemes, we obtain functions qualitatively similar to that shown in Fig.~\ref{fig:exclusionplot}. In physical volume weighting, the deviations from the Gaussian model arise at relatively smaller values of $A_\phi$ and larger values of $H_{\rm I} L_{\rm obs}$. This can be traced back to the onset of non-Gaussianities in $a_{20}^{\rm (UL)}$ at smaller values of $A_\phi$ for physical volume weighting (see Fig.~\ref{fig:quadrupoleplots}). For large-field and intermediate models of inflation, the onset of deviations from the Gaussian model arise at larger values of $A_\phi$ and smaller values of $H_{\rm I} L_{\rm obs}$ than for the small-field model. Therefore, while the precise constraints from the posterior depend on the inflationary potential and weighting scheme, the result in Fig.~\ref{fig:exclusionplot} is qualitatively robust.

\section{Extension to Three Dimensions}\label{sec:3d}
We now briefly outline how the approach above may be extended to three dimensions, but leave the use of three-dimensional numerical relativity simulations to verify the validity of these approximations to future work.
As above, we assume that $\zeta$ is the dominant gravitational degree of freedom.
As described in Appendix~\ref{app:loc}, in the Sachs-Wolfe approximation and to linear order in the local derivatives of $\zeta$ we have
\begin{equation}
  \label{eqn:a2m-3d}
  a^{\rm (UL)}_{2m} \propto -\left(H_{\rm I}L_{\rm obs}\right)^2e^{-2\sigma_\zeta \hat{G}_0}\hat{G}_{2m} \, ,
\end{equation}
where there are now $5$ different random deviates $\hat{G}_{2m}$ which arise as various linear combinations of the components of the Hessian of $\zeta$.
For statistically homogeneous and isotropic ULSS, we have $\langle \hat{G}_0\hat{G}_{2m}\rangle = 0$ and $\langle\hat{G}_{2m}\hat{G}_{2m'}\rangle \propto \delta_{m,m'}$.  
Nontrivial correlations between these variables arise from the breaking of statistical isotropy, an explicit example of which is given for the case of planar symmetric fluctuations above.
To full second order in the gradient expansion, an additional function of $\zeta$ and $(\partial_i\zeta\partial_j\zeta)^2$ (which is multilinear in the $\partial_i\zeta\partial_j\zeta$'s) can appear.
However, since this is a nonlinear correction in the local perturbation $\Delta\zeta$ (see~\eqref{eqn:local-zeta-pert}), it requires determining photon transport properties beyond linear perturbation theory, which we will not pursue here.\footnote{Such terms arise, for example, if the naive Sachs-Wolfe approximation is altered by evaluating $\zeta$ at the true physical distance to the last scattering surface, rather than the physical distance measured in the locally defined FRW background.  Furthermore, they are generated by the coordinate transform that removes the gradient term in the Taylor expansion of $\zeta$.}
The important feature for the analysis we presented above is the presence of the $e^{-2\zeta}$ multiplier, which leads to the emergence of a strong peak in the $a_{2m}$ distribution around zero as before. 
As well as the CMB quadrupole, we can estimate the locally measured value of $\Omega_k$ using~\eqref{eqn:three-ricci} and~\eqref{eqn:omega-k-ricci}:
\begin{equation}
  \label{eqn:omega-k}
  \Omega_k \approx -\frac{e^{-2\zeta}}{6H_0^2}\left(4\nabla^2\zeta + 2(\nabla\zeta)^2\right) \, ,
\end{equation}
where $\nabla^2\zeta$ is a Gaussian random deviate and $(\nabla\zeta)^2$ a $\chi^2$ deviate with three degrees of freedom in the Gaussian random field model for $\zeta$.

\begin{figure}
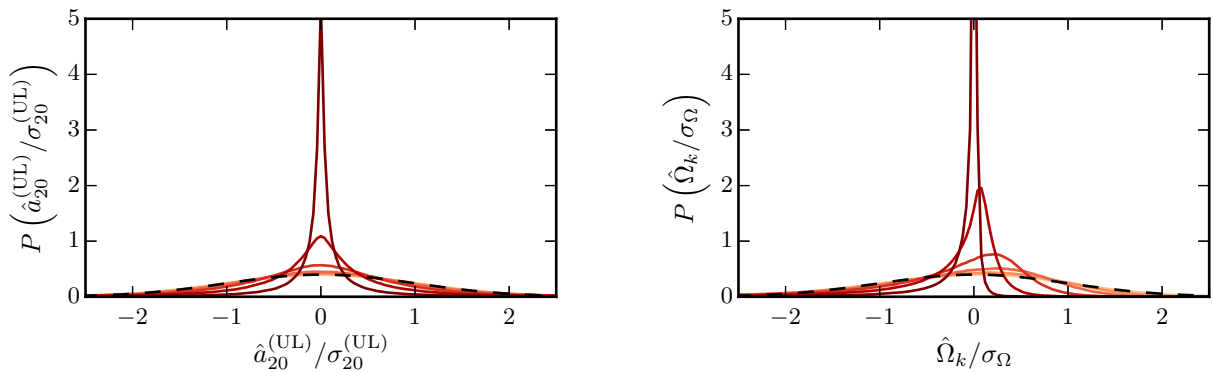

  \includegraphics[width=3.375in]{{{alm-vary-amp-3d-1-1p56}}}
  \includegraphics[width=3.375in]{{{ricci-vary-amp-1-1p56}}}
  \caption{\emph{Left:} Normalized distributions of $a_{2m}^{\rm (UL)}$ for our Gaussian random field model of $\zeta$.  The width of the distributions are normalized to $\left(\sigma_{2m}^{\rm (UL}\right)^2 \equiv \langle {a_{2m}^{\rm (UL)}}^2\rangle = e^{8\sigma_\zeta^2}\langle\hat{G}_{2m}^2\rangle$ where the averages weight each synchronous observer equally.  \emph{Right:} Normalized distributions of local values of $\Omega_k$ as estimated by~\eqref{eqn:omega-k}.  In each panel, the colors indicate the values of $\sigma_\zeta$, which are log uniformly distributed in the range $[10^{-2},0.7]$ with darker and redder lines corresponding to larger $\sigma_\zeta$.  In both figures, we have fixed $\left\langle \partial_{xx}\zeta^2\right\rangle / \sigma_\zeta = \frac{1.56}{5}$ and $\left\langle (\partial_x\zeta)^2\right\rangle / \sigma_\zeta = \frac{1}{3}$, resulting in $\langle(\nabla^2\zeta)^2\rangle / \sigma_\zeta = 1.56$ and $\left\langle \left(\nabla\zeta\right)^2\right\rangle / \sigma_\zeta = 1$.  The overall width of the Gaussian deviates $\hat{G}_{2m}$ scales out of the normalized distributions plotted here.}
  \label{fig:3D-distributions}
\end{figure}

In~\figref{fig:3D-distributions}, we show distributions for $a^{\rm (UL)}_{2m}$ generated from~\eqref{eqn:a2m-3d} (left panel) and for $\Omega_k$ from~\eqref{eqn:omega-k} (right panel), as we scan the variance in $\delta\zeta$ while holding $\langle (\partial_{i}^n\zeta)^2\rangle/\sigma_\zeta^2$ fixed.  These normalized variances can be related to various moments of the power spectrum through the three-dimensional generalization of~\eqref{eqn:shape-params}.  As in the case of planar symmetry, the distribution over $a^{\rm (UL)}_{2m}$ becomes sharply peaked around zero as the amplitude of fluctuations increases, with similar qualitative features also occurring in the distribution of the spatial curvature.
Therefore, we can predict that in the three dimensional scenario, measurements of the quadrupole and spatial curvature lead to relatively {\rm weaker constraints} on ULSS in the non-linear regime than would be expected by extrapolating the linear result.   

As well as the potential application to three-dimensions, the analysis in this section demonstrates the robustness of our observation that large amplitude initial fluctuations $\delta\phi$ in spatially flat gauge cannot be constrained by the CMB quadrupole.
In particular, the loss of constraining power in the planar symmetric scenario arises from the formation of the sharp peak in the $a_{20}$ distribution at $a_{20} = 0$.
However, we demonstrated this peak forms from the resulting large amplitude fluctuations in $\zeta$, which themselves arise from fluctuations in the initial location of the field on the potential.
For superhorizon perturbations, this correspondence between $\zeta$ and $\delta\phi$ fluctuations will be independent of the spatial dimension, and therefore the qualitative conclusions reached in this work should continue to hold in the higher dimensional setting.
Similarly, the choice of local volume weighting of observers will further decrease the constraining power of the quadrupole, as the regions that have inflated longer (and thus have pushed the local ULSS to the largest scales which are unobservable) will be preferentially weighted.

\section{Conclusions}
We have studied a toy model in which ultra-large scale structure (ULSS) is generated from the nonlinear evolution of large amplitude inhomogeneities in the initial conditions for inflation in a single spatial direction.  By enforcing $\mathcal{O}(60)$ $e$-folds of inflationary expansion in the homogeneous limit, we consider the most optimistic scenario for observable dynamically-evolved ULSS in single-field inflation.  In this setting, we have found a non-Gaussian probability distribution over the imprint of ULSS on the CMB quadrupole. We find that in cases where inflation is not completely disrupted, large amplitude pre-inflationary ULSS  is allowed by current data over a wide range in mapping scales $H_{\rm I} L_{\rm obs}$. The expectation from Gaussian statistics and linear cosmological perturbation theory is that increasingly large amplitude pre-inflationary ULSS requires an increasingly large number of inflationary $e$-folds, and hence a small mapping scale $H_{\rm I} L_{\rm obs}$, to not make a large contribution to the CMB quadrupole. Our results are at odds with this expectation, since increasingly large amplitude pre-inflationary ULSS is completely consistent with increasingly {\em large} mapping scales $H_{\rm I} L_{\rm obs}$. In this sense, inflation does a better job of hiding its initial conditions than you would expect based on linear theory and Gaussian statistics.

Qualitatively, based on a semi-analytic model, we have argued that these conclusions are expected to hold in a more realistic scenario where primordial inhomogeneities have no assumed symmetries. In a crude approximation, we expect that the contribution to $a_{2m}$ from ULSS for each $m$ is drawn from the non-Gaussian distributions similar to those in~\figref{fig:3D-distributions}, which are qualitatively similar to those of our toy model. The contribution from ULSS to the spatial curvature would be drawn from an analogously non-Gaussian distribution, making it less likely to observe spatial curvature than in previous computations that neglect gravitational nonlinearities~\cite{Vardanyan:2009ft,Kleban:2012ph}.
 
More generally, this work illustrates the novel possibility of applying the tools of numerical relativity to formulate precise tests of early universe physics. As we reach the limit on accessible information about the primordial universe, precise tools for making theoretical predictions are essential for extracting the most information from the limited observations that are possible to make.

\begin{acknowledgments}
JB and HVP are supported by the European Research Council under the European Community's Seventh Framework Programme (FP7/2007-2013) / ERC grant agreement no 306478-CosmicDawn. MCJ is supported by the National Science and Engineering Research Council through a Discovery grant. Research at Perimeter Institute is supported by the Government of Canada through Industry Canada and by the Province of Ontario through the Ministry of Research and Innovation. AA was supported in part by time release grant FQXi-PO-1501 from the Foundational Questions Institute (FQXi), of which he is Associate Director.  This work was supported in part by National Science Foundation Grant No. PHYS-1066293 and the hospitality of the Aspen Center for Physics. We also thank the Department of Physics at the University of Auckland for hospitality during the completion of this work. We thank R.\ Easther, D.\ Mortlock, and E.\ Lim for useful conversations.
\end{acknowledgments}

\appendix

\section{Numerical Approach}\label{app:numerical}
In this Appendix, we present the full evolution and constraint equations used in the main text, as well as a brief description of our numerical methods.
We focus on the evolution of a canonically normalized scalar field minimally coupled to gravity with action
\begin{equation}
  S = \int d^4 x \sqrt{-g} \left[ \frac{1}{2 M_{p}^2}  R + \mathcal{L}(\partial_\mu \phi, \phi) \right] \, ,
\end{equation}
where $R$ is the four dimensional Ricci scalar and $\mathcal{L}(\partial_\mu \phi, \phi)$ is the Lagrangian density for the inflaton
\begin{equation}\label{eq:lagrangian}
  \mathcal{L} = -\frac{1}{2}\partial_\mu\phi\partial^\mu\phi - V_0\left(1-e^{-\sqrt{\frac{2}{3\alpha}}\frac{\phi}{M_P}} \right)^2 \, .
  \end{equation}
Varying the action, and assuming a metric of the form
\begin{equation}
  ds^2 = -d\tau^2 + \sclp(x,\tau)^2 dx^2 + \sclt(x,\tau)^2(dy^2+dz^2)
\end{equation}
with two free functions $\sclp(x,\tau)$ and $\sclt(x,\tau)$, one obtains a coupled set of partial differential equations with mixed first and second time derivatives.
By introducing the components of the extrinsic curvature tensor ${K^x}_x (x,\tau)$ and ${K^y}_y(x,\tau)$,
it is possible to isolate a first-order in time system of evolution equations for the gravitational sector and the matter sector:
\begin{subequations}\label{eq:EOMS}
  \begin{align}
    \dot{\sclp} &= - \sclp {K^x}_x, \\
    \dot{\sclt} &= - \sclt {K^y}_y, \\
    \dot{{K^x}_x} &= \frac{\sclt'^2}{\sclt^2 \sclp^2} + {{K^x}_x}^2 - {{K^y}_y}^2 + \frac{\left( \Pi_\phi^2 - \phi'^2 \right)}{2\sclp^2 \mpl^2}, \\
    \dot{{K^y}_y} &= - \frac{\sclt'^2}{2 \sclt^2 \sclp^2} + \frac{3}{2}  {{K^y}_y}^2 - \frac{V(\phi)}{2 \mpl^2} + \frac{\left( \Pi_\phi^2 + \phi'^2 \right)}{4\sclp^2 \mpl^2}, \\
    \dot{\Pi}_\phi &= 2 {K^y}_y \Pi_\phi + \frac{1}{\sclp} \phi'' + \left( \frac{2\sclt'}{\sclp\sclt}  - \frac{\sclp'}{\sclt^2} \right) \phi' -  \sclt \partial_{\phi} V(\phi), \\
    \dot{\phi} &= \frac{\Pi_\phi}{\sclp} \, .
  \end{align}
\end{subequations}
There are also two constraint equations, the Hamiltonian ($\mathcal{H}$) and momentum ($\mathcal{P}$) constraints, given by:
\begin{subequations}\label{eqn:constraints}
  \begin{align}
    \mathcal{H} &= \frac{2\sclt\sclp'\sclt'-\sclp\sclt'^2-2\sclp\sclt\sclt''}{\sclp^3 \sclt^2} + 2 {K^x}_x {K^y}_y \nonumber \\ & \qquad +  {{K^y}_y}^2 - \mpl^{-2}\left( \frac{\phi'^2 + \Pi_\phi^2}{2\sclp^2} + V \right), \\
    \mathcal{P} &= {{K^y}_y}' - \frac{\sclt'}{\sclt} \left({K^x}_x - {K^y}_y \right) - \frac{\phi' \Pi_\phi}{2\sclp \mpl^2} \, .
  \end{align}
\end{subequations}
Both the Hamiltonian and momentum constraints should be zero for exact solutions to Einstein's equations. Numerically, we require that they remain as small as possible.

The evolution equations are solved numerically using a 10th order Gauss-Legendre time integrator~\cite{Butcher:1964,Braden:2014cra} and a Fourier collocation-based spatial discretization~\cite{opac-b1130608}.
As expected, the code displays exponential convergence with spatial resolution (once all spatial structure is resolved) and tenth order convergence with temporal resolution.
Adaptive time-stepping is used to capture the relevant dynamical time scale, which varies by orders of magnitude during a single run.
We also refine our grid as necessary to ensure we properly capture all spatial structure as the fields evolve in time.
Through this combination of precision numerical methods, we are able to evolve a single field realization through $60$ $e$-folds of inflation and several post-inflationary $e$-folds of field oscillations in order $1$-$10$ seconds (dependent on $A_\phi$) on a single core.
All dynamical fields are resolved to machine precision, with a similar accuracy in constraint preservation.
This numerical efficiency allows us to explore a range of amplitudes $A_\phi$ spanning between four and six orders of magnitude (dependent on the choice of $\alpha$), running 100 realizations of the Gaussian initial conditions at each amplitude.

\section{Extraction of Local Observables in the Local Linear Approximation}\label{app:loc}
In this Appendix we briefly outline an approach for extracting local CMB observables from numerically generated spacetimes possessing complex superhorizon structure.
For superhorizon perturbations in a synchronous coordinate system during inflation, we have to leading order in the gradient expansion (assuming general relativity)
\begin{equation}
  \label{eqn:synchronous-metric}
  ds^2 = -d\tau^2 + e^{2\psi({\bf x},\tau)}\bar{\gamma}_{ij}({\bf x}) \qquad \mathrm{det}\bar{\gamma}_{ij} = 1 \, .
\end{equation}
The time-independence of $\bar{\gamma}_{ij}$ arises from the rapid damping of the trace-free part of the extrinsic curvature, which drives temporal variations in $\bar{\gamma}_{ij}$.
Re-slicing our spacetime so that our time-coordinate is defined by surfaces of constant $H = -\frac{1}{3}{K_i}^i = \partial_\tau\psi$ and leaving the spatial coordinates unchanged ${\bf X}({\bf x}) = {\bf x}$, the metric in the new coordinate system is
\begin{equation}
  \label{eqn:constant-H}
  ds^2 = -\frac{1}{\dot{H}^2}dH^2 + \frac{2}{\dot{H}^2}\partial_iHdHdX^i + e^{2\zeta_{\bar{\gamma}}}\left(\bar{\gamma}_{ij} - e^{-2\zeta}\frac{\partial_iH\partial_jH}{\dot{H}^2} \right)dX^idX^j \, .
\end{equation}
In~\eqref{eqn:constant-H}, the space and time derivatives are with respect to the original synchronous coordinate system
\begin{align}
  {}' = \left(\frac{\partial}{\partial x}\right)_{\tau={\rm const}} \qquad \dot{{}} = \left(\frac{\partial}{\partial\tau}\right)_{x={\rm const}} \, .
\end{align}
This provides the definition of $\zeta_{\bar{\gamma}}({\bf x},H)$, which has the important virtue of being of the form $\zeta_{\bar{\gamma}}({\bf x},H) = \zeta_{\bar{\gamma},0}(H) + \delta\zeta_{\bar{\gamma}}({\bf x})$, when the wavelength of the fluctuations $\delta\zeta$ is larger than the horizon.
Since synchronous coordinates are not uniquely defined, we have indicated this through the ${}_{\bar{\gamma}}$ subscript on $\zeta_{\bar{\gamma}}$.
After a short transient, we verified numerically that this freezeout does indeed occur.

We now assume that $\zeta_{\bar{\gamma}}$ is the dominant gravitational degree of freedom, so that we can choose a synchronous coordinate system~\eqref{eqn:synchronous-metric} with $\bar{\gamma}_{ij} = \delta_{ij}$.  Denote $\zeta_{\bar{\gamma}}$ in this coordinate system by $\zeta$.
We also drop the subdominant correction $e^{-2\zeta_{\bar{\gamma}}}\frac{\partial_iH\partial_jH}{\dot{H}^2}$ correction.
All cosmological observables are then encoded in $\zeta$.
Since we consider only superhorizon perturbations, at each point ${\bf x_0}$ we want to approximate our spacetime as a local FRW.  We therefore Taylor expand
\begin{equation}
  \label{eqn:local-zeta-pert}
  \zeta({\bf x}_0+\delta{\bf x}) \approx \zeta({\bf x}_0) + \delta{\bf x}\cdot\nabla\zeta + \frac{\delta x_i\delta x_j}{2}\frac{\partial^2\zeta}{\partial x_i\partial x_j} + \mathcal{O}(\nabla^3) \equiv \zeta({\bf x}_0) + \Delta\zeta(\delta{\bf x}|{\bf x}_0) = \bar{\zeta}+\delta\zeta({\bf x}_0) + \Delta\zeta(\delta{\bf x}|{\bf x}_0) \, .
\end{equation}
Provided we restrict ourselves to positions within the local Hubble horizon, each term beyond the first one will be small, and we can treat them as linear perturbations to a locally defined background FRW spacetime with locally defined scale factor $a(H|{\bf x}_0) = e^{\zeta({\bf x}_0,H)}$.
Working to linear order in $\nabla\zeta$ and $\partial_{ij}\zeta$, the effect of $\nabla\zeta$ can be removed via a coordinate change, and we can use the Sachs-Wolfe approximation to obtain the angular dependence of the CMB temperature fluctuations
\begin{equation}
  \label{eqn:Sachs-Wolfe}
   \frac{\delta T}{T_{\rm CMB}}(\theta,\phi|{\bf x}_0) \approx -\frac{1}{5}\Delta\zeta(\delta {\bf x}_{\rm ls}({\bf x}_0),H_{\rm ls}|{\bf x}_0) \, ,
\end{equation}
where $\delta {\bf x}_{\rm ls}$ is the location of the last scattering surface and $H_{\rm ls}$ is the Hubble rate at last scattering.
This approximation is derived within linear perturbation theory around a homogeneous FRW spacetime, and therefore to be consistent we assume that $r_{\rm ls}^2 \equiv \sum_i \delta x_{i, {\rm ls}}^2$ has no angular dependence.
Importantly, $r_{\rm ls}$ does depend on the observer location ${\bf x}_0$ through the spatial dependence of $\zeta({\bf x}_0)$, which leads to substantial deviations from the expectations based on linear perturbation theory.
Beyond linear perturbation theory in each local FRW spacetime, one may expect that $r_{\rm ls}$ acquires angular dependence (for example, if we evaluate $\zeta$ at the true physical distance to the last scattering surface).
This induces additional corrections involving $(\Delta\zeta)^2$ and higher order gradients to our results below.
However, as we will show below, the dominant effect on the qualitative behaviour of $a_{2m}^{\rm (UL)}$ arises solely from the ${\bf x}_0$ dependence of $r_{\rm ls}$, and therefore these corrections will not remove the loss of constraining power for large amplitude initial fluctuations $A_\phi$.

Carrying through the calculation outlined above, we find to linear order in the synchronous derivatives of $\zeta$
\begin{equation}
  \label{eqn:a2m-full}
 a_{2m}^{\rm (UL)} ({\bf x}_0) \propto -e^{-2\delta\zeta({\bf x}_0)}\hat{G}_{2m} \propto -e^{-2\sigma_\zeta\hat{G}_0}\hat{G}_{2m} \, ,
\end{equation}
where $\hat{G}_0$ and $\hat{G}_{2m}$ are (possibly correlated) random deviates determined by the statistics of the underlying field $\zeta$. In particular, $\hat{G}_{2m}$ are given by various linear combinations of the trace-free part of the Hessian for $\zeta$ resulting from the projection onto spherical harmonics at fixed $r_{\rm ls}$. 
This expression is only correct to \emph{linear} order in the derivatives of $\zeta$, not to second order in the gradient expansion.  In particular, an extra function of $(\partial_x\zeta)^2$ can appear (even within the approximation $\bar{\gamma}_{ij} = \delta_{ij}$).  A proper calculation of this effect requires moving beyond the derivation of the Sachs-Wolfe effect in linear perturbation theory, and will be presented elsewhere.
Similarly, the local value of the intrinsic curvature of the constant $H$ slices is
\begin{equation}
  \label{eqn:three-ricci}
  {}^{(3)}R \approx -e^{-2\zeta}\left(4\nabla^2\zeta + 2(\nabla\zeta)^2 \right) \, .
\end{equation}
Assuming $\bar{\gamma}_{ij} = \delta_{ij} + \bar{h}_{ij}$ can be treated perturbatively in $\bar{h}_{ij}$,~\eqref{eqn:three-ricci} is correct to second order in spatial derivatives.
Rather than deal directly with the intrinsic curvature of the slices, it is convenient to instead consider
\begin{equation}
  \label{eqn:omega-k-ricci}
  \Omega_k = \frac{{}^{(3)}R}{6H_0^2}
\end{equation}
which is identical up to an overall constant.
Although not precisely the same, this quantity will be closely related to the quantity identified as spatial curvature in cosmological observations, as it directly enters into the Hamiltonian constraint, which is the inhomogeneous generalization of the Friedmann equation.


\begin{thebibliography}{39}
\expandafter\ifx\csname natexlab\endcsname\relax\def\natexlab#1{#1}\fi
\expandafter\ifx\csname bibnamefont\endcsname\relax
  \def\bibnamefont#1{#1}\fi
\expandafter\ifx\csname bibfnamefont\endcsname\relax
  \def\bibfnamefont#1{#1}\fi
\expandafter\ifx\csname citenamefont\endcsname\relax
  \def\citenamefont#1{#1}\fi
\expandafter\ifx\csname url\endcsname\relax
  \def\url#1{\texttt{#1}}\fi
\expandafter\ifx\csname urlprefix\endcsname\relax\def\urlprefix{URL }\fi
\providecommand{\bibinfo}[2]{#2}
\providecommand{\eprint}[2][]{\url{#2}}

\bibitem[{\citenamefont{Aguirre}(2008)}]{Aguirre:2007gy}
\bibinfo{author}{\bibfnamefont{A.}~\bibnamefont{Aguirre}}, in
  \emph{\bibinfo{booktitle}{Beyond the Big Bang}}
  (\bibinfo{publisher}{Springer}, \bibinfo{year}{2008}).

\bibitem[{\citenamefont{Guth}(2007)}]{Guth:2007ng}
\bibinfo{author}{\bibfnamefont{A.~H.} \bibnamefont{Guth}},
  \bibinfo{journal}{J.Phys.} \textbf{\bibinfo{volume}{A40}},
  \bibinfo{pages}{6811} (\bibinfo{year}{2007}), \eprint{hep-th/0702178}.

\bibitem[{\citenamefont{Grischuk and Zel'Dovich}(1978)}]{GZeffect}
\bibinfo{author}{\bibfnamefont{L.~P.} \bibnamefont{Grischuk}} \bibnamefont{and}
  \bibinfo{author}{\bibfnamefont{Y.~B.} \bibnamefont{Zel'Dovich}},
  \bibinfo{journal}{Astron. Zh.} \textbf{\bibinfo{volume}{55}},
  \bibinfo{pages}{209} (\bibinfo{year}{1978}).

\bibitem[{\citenamefont{Castro et~al.}(2003)\citenamefont{Castro, Douspis, and
  Ferreira}}]{Castro:2003bk}
\bibinfo{author}{\bibfnamefont{P.~G.} \bibnamefont{Castro}},
  \bibinfo{author}{\bibfnamefont{M.}~\bibnamefont{Douspis}}, \bibnamefont{and}
  \bibinfo{author}{\bibfnamefont{P.~G.} \bibnamefont{Ferreira}},
  \bibinfo{journal}{Phys. Rev.} \textbf{\bibinfo{volume}{D68}},
  \bibinfo{pages}{127301} (\bibinfo{year}{2003}), \eprint{astro-ph/0309320}.

\bibitem[{\citenamefont{Turner}(1991)}]{Turner:1991dn}
\bibinfo{author}{\bibfnamefont{M.~S.} \bibnamefont{Turner}},
  \bibinfo{journal}{Phys. Rev.} \textbf{\bibinfo{volume}{D44}},
  \bibinfo{pages}{3737} (\bibinfo{year}{1991}).

\bibitem[{\citenamefont{Goldwirth and Piran}(1990)}]{Goldwirth:1989pr}
\bibinfo{author}{\bibfnamefont{D.~S.} \bibnamefont{Goldwirth}}
  \bibnamefont{and} \bibinfo{author}{\bibfnamefont{T.}~\bibnamefont{Piran}},
  \bibinfo{journal}{Phys. Rev. Lett.} \textbf{\bibinfo{volume}{64}},
  \bibinfo{pages}{2852} (\bibinfo{year}{1990}).

\bibitem[{\citenamefont{Goldwirth}(1991)}]{Goldwirth:1990pm}
\bibinfo{author}{\bibfnamefont{D.~S.} \bibnamefont{Goldwirth}},
  \bibinfo{journal}{Phys.Rev.} \textbf{\bibinfo{volume}{D43}},
  \bibinfo{pages}{3204} (\bibinfo{year}{1991}).

\bibitem[{\citenamefont{Kurki-Suonio et~al.}(1993)\citenamefont{Kurki-Suonio,
  Laguna, and Matzner}}]{KurkiSuonio:1993fg}
\bibinfo{author}{\bibfnamefont{H.}~\bibnamefont{Kurki-Suonio}},
  \bibinfo{author}{\bibfnamefont{P.}~\bibnamefont{Laguna}}, \bibnamefont{and}
  \bibinfo{author}{\bibfnamefont{R.~A.} \bibnamefont{Matzner}},
  \bibinfo{journal}{Phys.Rev.} \textbf{\bibinfo{volume}{D48}},
  \bibinfo{pages}{3611} (\bibinfo{year}{1993}), \eprint{astro-ph/9306009}.

\bibitem[{\citenamefont{Johnson et~al.}(2012)\citenamefont{Johnson, Peiris, and
  Lehner}}]{Johnson:2011wt}
\bibinfo{author}{\bibfnamefont{M.~C.} \bibnamefont{Johnson}},
  \bibinfo{author}{\bibfnamefont{H.~V.} \bibnamefont{Peiris}},
  \bibnamefont{and} \bibinfo{author}{\bibfnamefont{L.}~\bibnamefont{Lehner}},
  \bibinfo{journal}{Phys. Rev.} \textbf{\bibinfo{volume}{D85}},
  \bibinfo{pages}{083516} (\bibinfo{year}{2012}), \eprint{1112.4487}.

\bibitem[{\citenamefont{Wainwright et~al.}(2014)\citenamefont{Wainwright,
  Johnson, Peiris, Aguirre, Lehner et~al.}}]{Wainwright:2013lea}
\bibinfo{author}{\bibfnamefont{C.~L.} \bibnamefont{Wainwright}},
  \bibinfo{author}{\bibfnamefont{M.~C.} \bibnamefont{Johnson}},
  \bibinfo{author}{\bibfnamefont{H.~V.} \bibnamefont{Peiris}},
  \bibinfo{author}{\bibfnamefont{A.}~\bibnamefont{Aguirre}},
  \bibinfo{author}{\bibfnamefont{L.}~\bibnamefont{Lehner}},
  \bibnamefont{et~al.}, \bibinfo{journal}{JCAP}
  \textbf{\bibinfo{volume}{1403}}, \bibinfo{pages}{030} (\bibinfo{year}{2014}),
  \eprint{1312.1357}.

\bibitem[{\citenamefont{East et~al.}(2015)\citenamefont{East, Kleban, Linde,
  and Senatore}}]{East:2015ggf}
\bibinfo{author}{\bibfnamefont{W.~E.} \bibnamefont{East}},
  \bibinfo{author}{\bibfnamefont{M.}~\bibnamefont{Kleban}},
  \bibinfo{author}{\bibfnamefont{A.}~\bibnamefont{Linde}}, \bibnamefont{and}
  \bibinfo{author}{\bibfnamefont{L.}~\bibnamefont{Senatore}}
  (\bibinfo{year}{2015}), \eprint{1511.05143}.

\bibitem[{\citenamefont{Giblin et~al.}(2016)\citenamefont{Giblin, Mertens, and
  Starkman}}]{Giblin:2015vwq}
\bibinfo{author}{\bibfnamefont{J.~T.} \bibnamefont{Giblin}},
  \bibinfo{author}{\bibfnamefont{J.~B.} \bibnamefont{Mertens}},
  \bibnamefont{and} \bibinfo{author}{\bibfnamefont{G.~D.}
  \bibnamefont{Starkman}}, \bibinfo{journal}{Phys. Rev. Lett.}
  \textbf{\bibinfo{volume}{116}}, \bibinfo{pages}{251301}
  (\bibinfo{year}{2016}), \eprint{1511.01105}.

\bibitem[{\citenamefont{Bentivegna and Bruni}(2016)}]{Bentivegna:2015flc}
\bibinfo{author}{\bibfnamefont{E.}~\bibnamefont{Bentivegna}} \bibnamefont{and}
  \bibinfo{author}{\bibfnamefont{M.}~\bibnamefont{Bruni}},
  \bibinfo{journal}{Phys. Rev. Lett.} \textbf{\bibinfo{volume}{116}},
  \bibinfo{pages}{251302} (\bibinfo{year}{2016}), \eprint{1511.05124}.

\bibitem[{\citenamefont{Mertens et~al.}(2016)\citenamefont{Mertens, Giblin, and
  Starkman}}]{Mertens:2015ttp}
\bibinfo{author}{\bibfnamefont{J.~B.} \bibnamefont{Mertens}},
  \bibinfo{author}{\bibfnamefont{J.~T.} \bibnamefont{Giblin}},
  \bibnamefont{and} \bibinfo{author}{\bibfnamefont{G.~D.}
  \bibnamefont{Starkman}}, \bibinfo{journal}{Phys. Rev.}
  \textbf{\bibinfo{volume}{D93}}, \bibinfo{pages}{124059}
  (\bibinfo{year}{2016}), \eprint{1511.01106}.

\bibitem[{\citenamefont{Adamek et~al.}(2016{\natexlab{a}})\citenamefont{Adamek,
  Daverio, Durrer, and Kunz}}]{Adamek:2015eda}
\bibinfo{author}{\bibfnamefont{J.}~\bibnamefont{Adamek}},
  \bibinfo{author}{\bibfnamefont{D.}~\bibnamefont{Daverio}},
  \bibinfo{author}{\bibfnamefont{R.}~\bibnamefont{Durrer}}, \bibnamefont{and}
  \bibinfo{author}{\bibfnamefont{M.}~\bibnamefont{Kunz}},
  \bibinfo{journal}{Nature Phys.} \textbf{\bibinfo{volume}{12}},
  \bibinfo{pages}{346} (\bibinfo{year}{2016}{\natexlab{a}}),
  \eprint{1509.01699}.

\bibitem[{\citenamefont{Kleban and Senatore}(2016)}]{Kleban:2016sqm}
\bibinfo{author}{\bibfnamefont{M.}~\bibnamefont{Kleban}} \bibnamefont{and}
  \bibinfo{author}{\bibfnamefont{L.}~\bibnamefont{Senatore}}
  (\bibinfo{year}{2016}), \eprint{1602.03520}.

\bibitem[{\citenamefont{Adamek et~al.}(2016{\natexlab{b}})\citenamefont{Adamek,
  Daverio, Durrer, and Kunz}}]{Adamek:2016zes}
\bibinfo{author}{\bibfnamefont{J.}~\bibnamefont{Adamek}},
  \bibinfo{author}{\bibfnamefont{D.}~\bibnamefont{Daverio}},
  \bibinfo{author}{\bibfnamefont{R.}~\bibnamefont{Durrer}}, \bibnamefont{and}
  \bibinfo{author}{\bibfnamefont{M.}~\bibnamefont{Kunz}},
  \bibinfo{journal}{JCAP} \textbf{\bibinfo{volume}{1607}}, \bibinfo{pages}{053}
  (\bibinfo{year}{2016}{\natexlab{b}}), \eprint{1604.06065}.

\bibitem[{\citenamefont{Clough et~al.}(2016)\citenamefont{Clough, Lim, DiNunno,
  Fischler, Flauger, and Paban}}]{Clough:2016ymm}
\bibinfo{author}{\bibfnamefont{K.}~\bibnamefont{Clough}},
  \bibinfo{author}{\bibfnamefont{E.~A.} \bibnamefont{Lim}},
  \bibinfo{author}{\bibfnamefont{B.~S.} \bibnamefont{DiNunno}},
  \bibinfo{author}{\bibfnamefont{W.}~\bibnamefont{Fischler}},
  \bibinfo{author}{\bibfnamefont{R.}~\bibnamefont{Flauger}}, \bibnamefont{and}
  \bibinfo{author}{\bibfnamefont{S.}~\bibnamefont{Paban}}
  (\bibinfo{year}{2016}), \eprint{1608.04408}.

\bibitem[{\citenamefont{Kallosh et~al.}(2013)\citenamefont{Kallosh, Linde, and
  Roest}}]{Kallosh:2013yoa}
\bibinfo{author}{\bibfnamefont{R.}~\bibnamefont{Kallosh}},
  \bibinfo{author}{\bibfnamefont{A.}~\bibnamefont{Linde}}, \bibnamefont{and}
  \bibinfo{author}{\bibfnamefont{D.}~\bibnamefont{Roest}},
  \bibinfo{journal}{JHEP} \textbf{\bibinfo{volume}{11}}, \bibinfo{pages}{198}
  (\bibinfo{year}{2013}), \eprint{1311.0472}.

\bibitem[{\citenamefont{Lyth}(1997)}]{Lyth:1996im}
\bibinfo{author}{\bibfnamefont{D.~H.} \bibnamefont{Lyth}},
  \bibinfo{journal}{Phys. Rev. Lett.} \textbf{\bibinfo{volume}{78}},
  \bibinfo{pages}{1861} (\bibinfo{year}{1997}), \eprint{hep-ph/9606387}.

\bibitem[{\citenamefont{Ade et~al.}(2015{\natexlab{a}})}]{Ade:2015tva}
\bibinfo{author}{\bibfnamefont{P.}~\bibnamefont{Ade}} \bibnamefont{et~al.}
  (\bibinfo{collaboration}{BICEP2, Planck}), \bibinfo{journal}{Phys. Rev.
  Lett.} \textbf{\bibinfo{volume}{114}}, \bibinfo{pages}{101301}
  (\bibinfo{year}{2015}{\natexlab{a}}), \eprint{1502.00612}.

\bibitem[{\citenamefont{Linde}(1983)}]{Linde:1983gd}
\bibinfo{author}{\bibfnamefont{A.~D.} \bibnamefont{Linde}},
  \bibinfo{journal}{Phys. Lett.} \textbf{\bibinfo{volume}{B129}},
  \bibinfo{pages}{177} (\bibinfo{year}{1983}).

\bibitem[{\citenamefont{Starobinsky}(1980)}]{Starobinsky:1980te}
\bibinfo{author}{\bibfnamefont{A.~A.} \bibnamefont{Starobinsky}},
  \bibinfo{journal}{Phys. Lett.} \textbf{\bibinfo{volume}{B91}},
  \bibinfo{pages}{99} (\bibinfo{year}{1980}).

\bibitem[{\citenamefont{Ade et~al.}(2015{\natexlab{b}})}]{Ade:2015xua}
\bibinfo{author}{\bibfnamefont{P.~A.~R.} \bibnamefont{Ade}}
  \bibnamefont{et~al.} (\bibinfo{collaboration}{Planck})
  (\bibinfo{year}{2015}{\natexlab{b}}), \eprint{1502.01589}.

\bibitem[{\citenamefont{Xue et~al.}(2013)\citenamefont{Xue, Garfinkle,
  Pretorius, and Steinhardt}}]{Xue:2013bva}
\bibinfo{author}{\bibfnamefont{B.}~\bibnamefont{Xue}},
  \bibinfo{author}{\bibfnamefont{D.}~\bibnamefont{Garfinkle}},
  \bibinfo{author}{\bibfnamefont{F.}~\bibnamefont{Pretorius}},
  \bibnamefont{and} \bibinfo{author}{\bibfnamefont{P.~J.}
  \bibnamefont{Steinhardt}}, \bibinfo{journal}{Phys.Rev.}
  \textbf{\bibinfo{volume}{D88}}, \bibinfo{pages}{083509}
  (\bibinfo{year}{2013}), \eprint{1308.3044}.

\bibitem[{\citenamefont{Johnson et~al.}(2015)\citenamefont{Johnson, Wainwright,
  Aguirre, and Peiris}}]{Johnson:2015gma}
\bibinfo{author}{\bibfnamefont{M.~C.} \bibnamefont{Johnson}},
  \bibinfo{author}{\bibfnamefont{C.~L.} \bibnamefont{Wainwright}},
  \bibinfo{author}{\bibfnamefont{A.}~\bibnamefont{Aguirre}}, \bibnamefont{and}
  \bibinfo{author}{\bibfnamefont{H.~V.} \bibnamefont{Peiris}}
  (\bibinfo{year}{2015}), \eprint{1508.03641}.

\bibitem[{\citenamefont{Salopek and Bond}(1990)}]{Salopek:1990jq}
\bibinfo{author}{\bibfnamefont{D.~S.} \bibnamefont{Salopek}} \bibnamefont{and}
  \bibinfo{author}{\bibfnamefont{J.~R.} \bibnamefont{Bond}},
  \bibinfo{journal}{Phys. Rev.} \textbf{\bibinfo{volume}{D42}},
  \bibinfo{pages}{3936} (\bibinfo{year}{1990}).

\bibitem[{\citenamefont{Sasaki and Stewart}(1996)}]{Sasaki:1995aw}
\bibinfo{author}{\bibfnamefont{M.}~\bibnamefont{Sasaki}} \bibnamefont{and}
  \bibinfo{author}{\bibfnamefont{E.~D.} \bibnamefont{Stewart}},
  \bibinfo{journal}{Prog.Theor.Phys.} \textbf{\bibinfo{volume}{95}},
  \bibinfo{pages}{71} (\bibinfo{year}{1996}), \eprint{astro-ph/9507001}.

\bibitem[{\citenamefont{Erickcek et~al.}(2008)\citenamefont{Erickcek, Carroll,
  and Kamionkowski}}]{Erickcek:2008jp}
\bibinfo{author}{\bibfnamefont{A.~L.} \bibnamefont{Erickcek}},
  \bibinfo{author}{\bibfnamefont{S.~M.} \bibnamefont{Carroll}},
  \bibnamefont{and}
  \bibinfo{author}{\bibfnamefont{M.}~\bibnamefont{Kamionkowski}},
  \bibinfo{journal}{Phys.Rev.} \textbf{\bibinfo{volume}{D78}},
  \bibinfo{pages}{083012} (\bibinfo{year}{2008}), \eprint{0808.1570}.

\bibitem[{\citenamefont{Aguirre et~al.}(2007)\citenamefont{Aguirre, Gratton,
  and Johnson}}]{Aguirre:2006ak}
\bibinfo{author}{\bibfnamefont{A.}~\bibnamefont{Aguirre}},
  \bibinfo{author}{\bibfnamefont{S.}~\bibnamefont{Gratton}}, \bibnamefont{and}
  \bibinfo{author}{\bibfnamefont{M.~C.} \bibnamefont{Johnson}},
  \bibinfo{journal}{Phys. Rev.} \textbf{\bibinfo{volume}{D75}},
  \bibinfo{pages}{123501} (\bibinfo{year}{2007}), \eprint{hep-th/0611221}.

\bibitem[{\citenamefont{Garriga et~al.}(2006)\citenamefont{Garriga,
  Schwartz-Perlov, Vilenkin, and Winitzki}}]{Garriga:2005av}
\bibinfo{author}{\bibfnamefont{J.}~\bibnamefont{Garriga}},
  \bibinfo{author}{\bibfnamefont{D.}~\bibnamefont{Schwartz-Perlov}},
  \bibinfo{author}{\bibfnamefont{A.}~\bibnamefont{Vilenkin}}, \bibnamefont{and}
  \bibinfo{author}{\bibfnamefont{S.}~\bibnamefont{Winitzki}},
  \bibinfo{journal}{JCAP} \textbf{\bibinfo{volume}{0601}}, \bibinfo{pages}{017}
  (\bibinfo{year}{2006}), \eprint{hep-th/0509184}.

\bibitem[{\citenamefont{Linde}(2007)}]{Linde:2006nw}
\bibinfo{author}{\bibfnamefont{A.~D.} \bibnamefont{Linde}},
  \bibinfo{journal}{JCAP} \textbf{\bibinfo{volume}{0701}}, \bibinfo{pages}{022}
  (\bibinfo{year}{2007}), \eprint{hep-th/0611043}.

\bibitem[{\citenamefont{Brustein and Steinhardt}(1993)}]{Brustein:1992nk}
\bibinfo{author}{\bibfnamefont{R.}~\bibnamefont{Brustein}} \bibnamefont{and}
  \bibinfo{author}{\bibfnamefont{P.~J.} \bibnamefont{Steinhardt}},
  \bibinfo{journal}{Phys. Lett.} \textbf{\bibinfo{volume}{B302}},
  \bibinfo{pages}{196} (\bibinfo{year}{1993}), \eprint{hep-th/9212049}.

\bibitem[{\citenamefont{{Hinshaw} et~al.}(1996)\citenamefont{{Hinshaw},
  {Banday}, {Bennett}, {Gorski}, {Kogut}, {Smoot}, and
  {Wright}}}]{1996ApJ...464L..17H}
\bibinfo{author}{\bibfnamefont{G.}~\bibnamefont{{Hinshaw}}},
  \bibinfo{author}{\bibfnamefont{A.~J.} \bibnamefont{{Banday}}},
  \bibinfo{author}{\bibfnamefont{C.~L.} \bibnamefont{{Bennett}}},
  \bibinfo{author}{\bibfnamefont{K.~M.} \bibnamefont{{Gorski}}},
  \bibinfo{author}{\bibfnamefont{A.}~\bibnamefont{{Kogut}}},
  \bibinfo{author}{\bibfnamefont{G.~F.} \bibnamefont{{Smoot}}},
  \bibnamefont{and} \bibinfo{author}{\bibfnamefont{E.~L.}
  \bibnamefont{{Wright}}}, \bibinfo{journal}{ApjL}
  \textbf{\bibinfo{volume}{464}}, \bibinfo{pages}{L17} (\bibinfo{year}{1996}),
  \eprint{astro-ph/9601058}.

\bibitem[{\citenamefont{Vardanyan et~al.}(2009)\citenamefont{Vardanyan, Trotta,
  and Silk}}]{Vardanyan:2009ft}
\bibinfo{author}{\bibfnamefont{M.}~\bibnamefont{Vardanyan}},
  \bibinfo{author}{\bibfnamefont{R.}~\bibnamefont{Trotta}}, \bibnamefont{and}
  \bibinfo{author}{\bibfnamefont{J.}~\bibnamefont{Silk}},
  \bibinfo{journal}{Mon. Not. Roy. Astron. Soc.}
  \textbf{\bibinfo{volume}{397}}, \bibinfo{pages}{431} (\bibinfo{year}{2009}),
  \eprint{0901.3354}.

\bibitem[{\citenamefont{Kleban and Schillo}(2012)}]{Kleban:2012ph}
\bibinfo{author}{\bibfnamefont{M.}~\bibnamefont{Kleban}} \bibnamefont{and}
  \bibinfo{author}{\bibfnamefont{M.}~\bibnamefont{Schillo}},
  \bibinfo{journal}{JCAP} \textbf{\bibinfo{volume}{1206}}, \bibinfo{pages}{029}
  (\bibinfo{year}{2012}), \eprint{1202.5037}.

\bibitem[{\citenamefont{Butcher}(1964)}]{Butcher:1964}
\bibinfo{author}{\bibfnamefont{J.}~\bibnamefont{Butcher}},
  \bibinfo{journal}{Math. Comp.} \textbf{\bibinfo{volume}{18}},
  \bibinfo{pages}{50} (\bibinfo{year}{1964}).

\bibitem[{\citenamefont{Braden et~al.}(2015)\citenamefont{Braden, Bond, and
  Mersini-Houghton}}]{Braden:2014cra}
\bibinfo{author}{\bibfnamefont{J.}~\bibnamefont{Braden}},
  \bibinfo{author}{\bibfnamefont{J.~R.} \bibnamefont{Bond}}, \bibnamefont{and}
  \bibinfo{author}{\bibfnamefont{L.}~\bibnamefont{Mersini-Houghton}},
  \bibinfo{journal}{JCAP} \textbf{\bibinfo{volume}{1503}}, \bibinfo{pages}{007}
  (\bibinfo{year}{2015}), \eprint{1412.5591}.

\bibitem[{\citenamefont{Boyd}(2001)}]{opac-b1130608}
\bibinfo{author}{\bibfnamefont{J.~P.} \bibnamefont{Boyd}},
  \emph{\bibinfo{title}{Chebyshev and Fourier spectral methods}}
  (\bibinfo{publisher}{Dover Publ. cop.}, \bibinfo{year}{2001}).

\end{thebibliography}
\end{document}